\documentclass[useAMS]{mn2e}

\usepackage{latexsym,graphicx}

\newcommand\kms{{\rm\,km\,s^{-1}}}

\newcommand\msun{\rm\,M_\odot}
\newcommand\rsun{\rm\,R_\odot}
\newcommand\mbh{M_{\rm BH}}

\def\apgt{\ {\raise-.5ex\hbox{$\buildrel>\over\sim$}}\ }
\def\aplt{\ {\raise-.5ex\hbox{$\buildrel<\over\sim$}}\ }

\title[Hyperfast pulsars]{Hyperfast pulsars as the remnants of massive stars
ejected from young star clusters}
\author[V.V.Gvaramadze, A.Gualandris and S.Portegies Zwart]
       {Vasilii V. Gvaramadze$^{1}$\thanks{E-mail: vgvaram@mx.iki.rssi.ru},
       Alessia Gualandris$^{2}$\thanks{E-mail: alessiag@astro.rit.edu}
       and Simon Portegies Zwart$^{3}$\thanks{E-mail: spz@science.uva.nl}\\
       $^{1}$Sternberg Astronomical Institute, Moscow State University,
       Universitetskij Pr. 13, Moscow 119992, Russia\\
       $^{2}$Center for Computational Relativity and Gravitation,
       Rochester Institute of Technology, 85 Lomb Memorial Drive,\\ Rochester, NY 14623, USA\\
       $^{3}$Astronomical Institute `Anton Pannekoek' and Section
            Computational Science, University of Amsterdam,\\ Kruislaan 403,
            1098 SJ, Amsterdam, the Netherlands}
\begin{document}

\date{Accepted 2007 December 19. Received 2007 December 11; in original form 2007 February 27}


\maketitle

\begin{abstract}
Recent proper motion and parallax measurements for the pulsar PSR
B1508+55 indicate a transverse velocity of $\sim 1\,100 \, {\rm km} \,
{\rm s}^{-1}$, which exceeds earlier measurements for any neutron star.
The spin-down characteristics of PSR
B1508+55 are typical for a non-recycled pulsar, which implies that the
velocity of the pulsar cannot have originated from the second
supernova disruption of a massive binary system. The high velocity of
PSR B1508+55 can be accounted for by assuming that it received a kick
at birth or that the neutron star was accelerated after its formation
in the supernova explosion.
We propose an explanation for the origin of hyperfast neutron stars based
on the hypothesis that they could be the remnants of a symmetric
supernova explosion of a high-velocity massive star which attained its
peculiar velocity (similar to that of the pulsar) in the course of a
strong dynamical three- or four-body encounter in the core of dense young
star cluster. To check this hypothesis we investigated three dynamical
processes involving close encounters between: (i) two hard massive
binaries, (ii) a hard binary and an intermediate-mass black hole, and
(iii) a single star and a hard binary intermediate-mass black hole.
We find that main-sequence O-type stars cannot be ejected from young massive
star clusters with peculiar velocities high enough to explain the origin
of hyperfast neutron stars, but lower mass main-sequence stars or the
stripped helium cores of massive stars could be accelerated to
hypervelocities. Our explanation for the origin of hyperfast pulsars requires
a very dense stellar environment of the order of $10^6 -10^7 \, {\rm stars} \,
{\rm pc}^{-3}$). Although such high densities may exist during the core
collapse of young massive star clusters, we caution that they have never been
observed.
\end{abstract}

\begin{keywords}
stellar dynamics -- methods: N-body simulations -- stars: individual: RX J0822-4300
-- stars: neutron -- pulsars: general -- pulsars: individual: B1508+55.
\end{keywords}

\section{Introduction}
\label{sec:intro}

It has been known for a long time that the typical peculiar space
velocities of pulsars are an order of magnitude larger that those of
their progenitors -- the massive stars (e.g. Gott, Gunn \& Ostriker
1970). Subsequent proper motion measurements for pulsars suggested
that some of them could be very fast objects, with peculiar speeds of
up to $\sim 1\,000 \, {\rm km} \, {\rm s}^{-1}$ (e.g. Chatterjee \&
Cordes 2004; Hobbs et al. 2005). Two mechanisms have been proposed for
the origin of high-velocity pulsars. The first one involves the
disruption of tight (semidetached) massive binary systems following
the second (symmetric) supernova explosion (e.g. Blaauw 1961;
Iben \& Tutukov 1996). This mechanism, however, cannot produce
velocities in excess of $\sim 700 \, {\rm km} \, {\rm s}^{-1}$
(Portegies Zwart \& van den Heuvel 1999). Moreover, the high-velocity
pulsars formed in this way are the remnants of the first supernova
explosion and therefore should
be {\it recycled}, i.e. their spin characteristics should be
significantly affected by the stellar wind of the companion star --
the progenitor of the second supernova (cf. Chatterjee et al. 2005).
The second mechanism relies on a natal kick (caused by the asymmetry
of the supernova explosion; e.g. Shklovskii 1970; Dewey \& Cordes
1987) or on a post-natal acceleration (caused by the asymmetric
electromagnetic or neutrino emission from the newborn pulsar; e.g.
Tademaru \& Harrison 1975; Chugai 1984). In principle, this mechanism
can produce solitary high-velocity pulsars with ordinary spin-down
characteristics, i.e. non-recycled pulsars (e.g. Scheck et
al. 2006). Thus, detection of non-recycled pulsars moving with a
velocity $\apgt 1\,000 \, {\rm km} \, {\rm s}^{-1}$ would suggest that
the mechanism involving the binary disruption by the supernova
explosion does not apply to all cases. In the following we refer to
pulsars (neutron stars) moving with velocities of $\ga 1\,000 \, {\rm
km} \, {\rm s}^{-1}$ as `hyperfast pulsars' (a term introduced by
Chatterjee et al. 2005).

Until recently the best candidates for hyperfast pulsars were PSR
B2224+65 (associated with the well-known Guitar Nebula; Chatterjee
\& Cordes 2004) and PSR B2011+38 (see Hobbs et al. 2005 and
references therein), whose peculiar velocities (both $\sim 1\,500 \,
{\rm km} \, {\rm s}^{-1}$) were inferred on the basis of proper
motion measurements and dispersion measure distance estimates. The
uncertainties associated with the distance estimates, however, leave
a possibility that these velocities are overestimated. Recent proper
motion and parallax measurements for the pulsar PSR B1508+55 gave
the first example of a high velocity ($1\,083_{-90} ^{+103} \, {\rm
km} \, {\rm s}^{-1}$) directly measured for a (non-recycled) pulsar
(Chatterjee et al. 2005). This result proved the existence of a
population of hyperfast neutron stars, and implies that the peculiar
velocity of PSR B1508+55 cannot be solely due to the disruption of a
tight massive binary system. A possible way to account for the high
velocity is to assume that at least part of this velocity is due
to a natal kick or a post-natal acceleration (Chatterjee et al.
2005).

In this paper, we propose an alternative explanation for the origin
of hyperfast pulsars (cf. Gvaramadze 2007). We suggest that PSR
B1508+55 (as well as other hyperfast neutron stars\footnote{Another
example of a hyperfast neutron star was recently reported by Hui \&
Becker (2006) and Winkler \& Petre (2007). Their proper motion
measurements for the central compact object RX J0822-4300 in the
supernova remnant (SNR) Puppis\,A suggest that the peculiar velocity
of this neutron star could be as large as $\sim 1\,000-1\,500 \,
{\rm km} \,{\rm s}^{-1}$, provided that the distance to these
objects is $\sim 2$ kpc.}) could be the remnant of a symmetric
supernova explosion of a high-velocity massive star (or its helium
core) which attained its peculiar velocity (similar to that of the
pulsar) in the course of a strong three- or four-body dynamical
encounter in the core of the parent young massive star cluster
(YMSC).

Our suggestion is based on the fact that massive
($\geq 10^4\msun$) star clusters are still forming in the disk of the
Milky Way (see Sect.\,\ref{sec:ymsc}) and on the hypothesis that the
cores of some YMSCs harbour intermediate-mass black holes (IMBHs),
i.e. black holes (BHs) with masses ranging from $\sim 100$ to $\sim
10^4\msun$ (see Sect.\,\ref{sec:ymsc}). Our suggestion also requires
the existence of very dense stellar environment ($\sim 10^6 -10^7 \,
{\rm pc}^{-3}$) to ensure that three- or four-body dynamical encounters
are frequent (see Sect.\,\ref{sec:hfp}). Although such high densities
may exist during the core collapse of YMSCs, we caution
that they have never been observed (we discuss this problem in
Sect.\,\ref{sec:ymsc}). In Sect.\,\ref{sec:hvs}, we briefly discuss the
mechanisms for the origin of hypervelocity stars,
the recently discovered class of stars moving with a speed of $\sim
1\,000 \, {\rm km} \,{\rm s}^{-1}$, and suggest that similar
mechanisms (although acting in a different environment) could be
responsible for the origin of some of the hyperfast neutron stars. In
Sect.\,\ref{sec:hfp}, we estimate the maximum possible velocities of
high-velocity escapers produced by various dynamical processes in the
cores of YMSCs. In Sect.\,\ref{sec:sim}, we compare these estimates
with results from numerical simulations while in Sect.\,\ref{sec:disc}
we discuss the ability of these processes to explain the origin of
hyperfast pulsars.

\section{Hypervelocity stars and mechanisms for their production}
\label{sec:hvs}

The discovery of high velocity pulsar PSR B1508+55 (Chatterjee et
al. 2005) coincided with the discovery of the hypervelocity
stars\footnote{In the following we use the term `hypervelocity'
[introduced by Hills (1988)] to designate the ordinary stars moving
with a peculiar speed of $\ga 1\,000 \, {\rm km} \,{\rm s}^{-1}$ while
for neutron stars we reserve the term `hyperfast' (see Sect.\,1).}
(Brown et al. 2005; Edelmann et al 2005; Hirsch et al. 2005). The
existence of the latter was predicted by Hills (1988), who showed that
a close encounter between a tight binary system and the supermassive
BH in the Galactic centre (e.g. Ghez et al.  2003; Sch\"{o}del et al.
2003) could be responsible for the ejection of one of the binary
components with a velocity of up to several $1\,000 \, {\rm km} \,
{\rm s}^{-1}$ (see also Hills 1991). Yu \& Tremaine (2003) proposed
two additional possible mechanisms for the production of hypervelocity
stars. The first one involves close encounters between two single
stars in the vicinity of a supermassive BH (the probability of this
process, however, is very low and we will not discuss it further)
while the second one is based on the interaction between a single star
and a putative binary BH in the Galactic centre (e.g. Hansen \&
Milosavljevi\'{c} 2003).

The discovery of hypervelocity stars (Brown et al. 2005;
Edelmann et al. 2005; Hirsch et al. 2005; Brown et al. 2006)
provides support for models of ejection mechanisms involving
dynamical processes in the vicinity of the supermassive BH
(Gualandris, Portegies Zwart \& Sipior 2005; Baumgardt, Gualandris
\& Portegies Zwart 2006; Bromley et al.  2006), and now it is widely
believed that these high-velocity objects originate in the Galactic
Centre (see, however, Edelmann et al. 2005; see also
Sect.\,\ref{sec:disc}). It is, therefore, possible that some
hyperfast pulsars (or their progenitor stars) were also ejected from
the Galactic Centre. For example, one of the hypervelocity stars is
a early B-type star of mass $\sim 8\msun$ (Edelmann et al. 2005), so
that this star will end its evolution in a Type II supernova leading
to the production of a hyperfast neutron star.
The proper motion measured for PSR B1508+55, however, indicates that
this pulsar was born in the Galactic disk near the Cyg OB
associations (Chatterjee et al. 2005), i.e. its origin cannot be
associated with the Galactic Centre\footnote{For a nice picture
illustrating the trajectory of the pulsar on the sky see {\tt
http://www.jb.man.ac.uk/news/fastestpulsar.html}}. The proper
motions of PSR B2224+65 and PSR B2011+38 also suggest that these
pulsars were born far from the Galactic Centre. Therefore, one should
look for other places in our Galaxy where one can find a
sufficiently massive BH and where the number density of the local
stellar population is high enough to ensure that close encounters
between stars and the BH are frequent. Note that the characteristic
age of the above mentioned pulsars (ranging from $\sim 0.4$ to $\sim
2.3$ Myr) implies that at least these three pulsars were not ejected
from the dense cores of globular clusters (whose typical age is
$\ga$ Gyr) and point to the more plausible cites of their origin --
the YMSCs.

We note that with the YMSCs could be associated another important
channel for production of high-velocity stars, namely through
close dynamical encounters between stars in their cores. This
process constitutes the base of the dynamical-ejection scenario
proposed by Poveda, Ruiz \& Allen (1967) to explain the origin of runaway
OB stars. The most effective path for production of high-velocity
stars by stellar encounters is through interaction between two
hard binaries (e.g. Mikkola 1983; Leonard \& Duncan 1988, 1990). This
process and processes involving dynamical encounters with IMBHs are
discussed in detail in Sect.\,\ref{sec:hfp}.

\section{Young massive star clusters}
\label{sec:ymsc}

Recent discovery of young ($\leq 10^7$ yr) and massive ($\geq 10^4\msun$)
star clusters (Borissova et al. 2006; Figer et al. 2006; Davies et al. 2007)
accompanied by upward revision of masses of the already known star clusters
(Kn\"{o}dlseder 2000; Clark et al. 2005; Homeier \& Alves 2005; Santos, Bonatto
\& Bica 2005; Ascenso et al. 2007; Wolff et al. 2007; Harayama, Eisenhauer \&
Martins 2008) increased the number of YMSCs in the Galactic disk to $\ga 10$.
All but one of these YMSCs
are located on the near side of the Galaxy. Taken together these facts suggest
that YMSCs are more numerous than it was known hitherto and that many of them
are still hidden from observers by the obscuring material in the Galactic plane
(Kn\"{o}dlseder et al. 2002; Hanson 2003).

Let us estimate the number of YMSCs formed in the Galactic disk during the last
$3\times 10^7$ yr (the lifetime of a star with mass of $8\,\msun$ -- the
minimum mass of single stars producing neutron stars). We choose this time-span
since we assume that the currently observed hyperfast neutron stars could be the
descendants of either a $8\, \msun$ hypervelocity star, ejected at the very
beginning of its life, or a stripped helium core of the more massive star,
attained its peculiar velocity only recently (see Sect.\,\ref{sec:hfp}).
Assuming that the star formation rate (SFR) of $\sim 7-10\times 10^{-4} \,
\msun \, {\rm yr}^{-1} {\rm kpc}^{-2}$, derived by Lada \& Lada (2003) for
embedded star clusters in the solar neighbourhood, is representative for the
Galactic disk as a whole (cf. Larsen 2006), one has that $\simeq 6.6-9.4\times
10^6 \, \msun$ were formed within a circle of 10 kpc. For a power-law cluster
initial mass function with a slope =2 (e.g. Lada \& Lada 2003) and $50 \, \msun
< M_{\rm cl} < 10^6 \, \msun$, one has $\sim 70-100$ YMSCs with a mass $M_{\rm cl}
\geq 10^4 \, M_{\odot}$. This estimate is in good agreement with that derived by
Kn\"{o}dlseder et al. (2002) using different arguments (see also Madhusudhan et
al. 2007), and exceeds by a factor of 2-3 the figures derived by Larsen (2006).
The discrepancy with Larsen (2006) originates due to the somewhat different input
parameters adopted in his paper. For example, he uses a factor of 1.3-1.9 smaller
SFR, derived by Lamers et al. (2006) for {\it bound} star clusters in the solar
neighbourhood, and assumes a larger upper cut-off of cluster masses, $M_{\rm cl}
<10^7 \, \msun$. We prefere to use the SFR given in Lada \& Lada (2003) to account
for the fact that many embedded star clusters independently of their mass became
dissolved within the first $\sim 10^7$ yr of their evolution (a possible example
of such a stellar system is the Cyg OB2 association).

To the YMSCs resided in the disk one should add the ones formed in the central
part of the Galaxy. Theoretical models of evolution and observability of star
clusters within 200 pc of the Galactic Centre suggest that this region could easily
harbour $\sim 50$ YMSCs with properties similar to those of the well-known Arches
and Quintuplet systems (Portegies Zwart et al. 2001a, 2002). Recent discovery with
2MASS of a large number of cluster candidates in direction to the Galactic Centre
(Dutra et al. 2003, and references therein) supports this suggestion, although
follow-up work is required to confirm the nature of these objects.

The dynamical evolution of YMSCs is dominated by massive stars, which
sink to the cluster centre on a time scale $t_{\rm cc} \sim 0.1-0.2 \,
t_{\rm rh}$, which typically is smaller than the main-sequence
lifetime of these stars. Interestingly, the dynamical friction time-scale
is independent of cluster mass, its size and density profile
(e.g. Portegies Zwart \& McMillan 2002; G\"urkan, Freitag \& Rasio
2004), where $t_{\rm rh}$ is the half-mass relaxation time-scale given
by (Spitzer 1987)
\begin{equation}
t_{\rm rh} \, \simeq \, {0.14 M_{\rm cl} ^{1/2} r_{\rm h} ^{3/2}
\over G^{1/2} \langle m \rangle \ln \Lambda }  \, . \label{relax}
\end{equation}
Here, $M_{\rm cl}$ and $r_{\rm h}$ are the total mass and the
characteristic (half-mass) radius of the cluster, $G$ is the
gravitational constant, $\langle m \rangle = M_{\rm cl} /N_{\rm cl}$
is the mean stellar mass, $N_{\rm cl}$ is the number of stars in the
cluster and $\ln \Lambda \simeq 10$ is the Coulomb logarithm. For
$t_{\rm cc} =0.15 \, t_{\rm rh}$ and using Eq.\,(\ref{relax}), one
has
\begin{eqnarray}
t_{\rm cc} \simeq 3\times 10^6 \, {\rm yr} \left({M_{\rm cl}
\over 10^4\msun}\right)^{1/2} \left({r_{\rm h} \over 1\,{\rm
pc} }\right)^{3/2} \ \nonumber \\
\times  \left({\langle m \rangle \over \msun}\right)^{-1} \left({\ln
\Lambda \over 10}\right)^{-1} \, . \label{cc}
\end{eqnarray}

As a result, mass segregation drives YMSCs to core collapse.  The
consequential increase in the central density may results in runaway
stellar collisions during which a massive star continues to grow in
mass due to repeated bombardment to a very massive star (VMS)
(Portegies Zwart et al. 1999; Portegies Zwart \& McMillan 2002;
G\"{u}rkan et al. 2004), provided that core collapse is completed
before the most massive stars in the cluster explode as supernovae,
i.e. $t_{\rm cc} \la 3\times 10^6$ yr. On this time-scale essentially
all massive ($\ga 20\msun$) stars in YMSCs segregate to the cluster
core. For $M_{\rm cl} = 10^4 - 10^5\msun$ and assuming a
$0.2-120\msun$ Salpeter initial mass function (in this case $\langle m
\rangle \simeq 0.7\msun$ and $N_{\rm cl} \simeq 1.5\times 10^4
-10^5$), one has from Eq.\,(\ref{cc}) that the runaway process occurs
in YMSCs with $r_{\rm h} \la 0.3-0.8\,$pc. The majority of young and
massive stars appear to reside in dense embedded clusters with a
characteristic radius of $\la 1$ pc, which is independent of mass
(Kroupa \& Boily 2002). It is therefore not unconceivable that an
appreciable fraction of YMSCs evolve through a collisional stage.
The maximum observed density in the cores of YMSCs [e.g. NGC\,3603
(Harayama et al. 2007), Arches (Figer et al. 1999)] is smaller than
that required for runaway collisions to occur\footnote{Note, however,
that the Arches cluster is not in a state of core collapse, but rather
on its way towards this (Portegies Zwart et al. 2007).}. Whether or not
the discrepancy between the observations and the theory is the result
of the small number of known YMSCs remains unclear to date. The
consequences of the collisional stage are profound, and we assume here
that it may exist at least in some YMSCs.

The VMS formed through runaway collisions will ultimately collapse to
a BH. One of the main questions here is wheather or not the BH will be
massive or not. The mass of the IMBH formed in this way could be as large
as $\sim 1\,000\msun$ for $M_{\rm cl} \sim 10^5\msun$ (Portegies Zwart
et al. 2004), but if the mass loss in the collision product is as high
as argued in some calculations (Belkus, Van Bever \& Vanbeveren, 2007;
Yungelson 2007; Yungelson et al 2008) the final BH may be $\la 100\,\msun$.

It is also possible that IMBHs of mass $\sim 10^3 -10^4\msun$ are the
descendants of old globular clusters disrupted by the tidal field of
the Galactic disk or that they are formed from the core collapse of
massive population III stars. Later they could be captured
gravitationally by molecular clouds or by the already existing young
star clusters. In the first case, the IMBH initiates star formation in
the cloud and, therefore, resides at the centre of the newly formed
star cluster (Miller \& Hamilton 2002), while in the second case the
IMBH rapidly sinks to the centre of the cluster due to the dynamical
friction (Miller \& Colbert 2004).

\section{Origin of hyperfast pulsars}
\label{sec:hfp}

We now consider the possibility that dynamical processes in the
cores of YMSCs could be responsible for ejection of hypervelocity
massive stars or their helium cores, whose subsequent collapse and
(symmetric) supernova explosion result in the origin of hyperfast
neutron stars (pulsars). In the following subsections, we estimate
the upper limits for the ejection speed produced by encounters
involving: (i) two hard binaries, (ii) a hard binary and
an IMBH and (iii) a single star and a hard binary IMBH.

\subsection{Binary-binary encounters}
\label{subsec:bb}

In Sect.\,\ref{sec:hvs}, we mentioned that the most effective path
for production of high-velocity stars by stellar encounters is
through interaction between two hard binaries, either the tidally
captured or the primordial ones. Note that the tidal binaries
are very hard by definition since the semimajor axes of their
circularized orbits are at most a few stellar radii (e.g. Lee \&
Ostriker 1986; McMillan, McDermott \& Taam 1987).
Therefore, it is likely that
collisions involving tidal binaries play a dominant role in
production of the fastest runaway stars. It is also important to
note that due to the mass segregation, the cores of YMSCs are
over-represented by massive stars, so that most of tidally captured
binaries should be the massive ones. Moreover, one can expect that
almost all tidal binaries would be massive if the most massive stars
are formed near the centres of their parent clusters (e.g. Bonnell,
Bate \& Zinnicker 1998; cf. Kroupa 2001).

Numerical simulations by Leonard \& Duncan (1988, 1990) showed
that the typical velocities at infinity (i.e. well after ejection)
of runaway stars (produced in the course
of binary-binary collisions) are similar to the orbital velocities
of the binary components while the velocities of some escapers can
be twice as large. Moreover, the maximum possible velocity
attained by the lightest member of the binaries involved in the
interaction (e.g. the helium core of a massive star or a early-type
B star) can be as large as the escape velocity from the
surface of the most massive star in the binaries (Leonard 1991).
For the upper main-sequence stars with the mass-radius
relationship (Habets \& Heintze 1981)
\begin{equation}
\label{eq:rms}
r_{\rm MS} = 0.8 \left({m_{\rm MS} \over \msun} \right)^{0.7}\rsun \, ,
\label{MS}
\end{equation}
where $r_{\rm MS}$ and $m_{\rm MS}$ are the stellar radius and
mass, the maximum possible velocity at infinity of ejected stars is a weak
function of $m_{\rm MS}$, $V_{{\infty},{\rm MS}} ^{\rm max} \simeq
700 \, {\rm km} \, {\rm s}^{-1} (m_{\rm MS} /\msun )^{0.15}$
and could be as large as $\sim 1\,400 \, {\rm km} \, {\rm s}^{-1}$
(cf. Leonard 1991). The ejection velocity could be even larger if
the binaries involved in the interaction are already evolved
through a common-envelope phase and consist of two helium cores.
In this case,
\begin{equation}
\label{eq:rhe}
r_{\rm He} \simeq 0.2 \left({m_{\rm He} \over \msun}
\right)^{0.65} \rsun \, , \label{He}
\end{equation}
where $r_{\rm He}$ and $m_{\rm He}$ are the radius and the mass of
a helium core (Tauris \& van den Heuvel 2006), so that
$V_{{\infty},{\rm He}} ^{\rm max} \simeq 1\,400 \, {\rm km} \, {\rm s}^{-1}
(m_{\rm He} /\msun )^{0.175} \sim 2\,300 \, {\rm km} \, {\rm
s}^{-1}$.

Scattering experiments by Leonard (1991) showed that only a small
fraction (less than 1 per cent) of runaway stars released in the
course of binary-binary collisions can attain the maximum possible
velocity. So that, to produce at least one hypervelocity (massive)
star during the first several Myr of cluster evolution,
the collisional time-scale for binaries in the cluster core should
be $\la 10^4$ yr. The total rate of binary-binary
encounters in the core of radius $r_{\rm c}$ is
\begin{equation}
\Gamma \, \sim \, {1\over 2} N_{\rm b} n_{\rm b} S_{\rm bb} V_{\rm rel}
\, , \label{rate}
\end{equation}
where $N_{\rm b} \simeq (4\pi /3)r_{\rm c} ^3 n_{\rm b}$, $n_{\rm
b}$ is the number density of binaries, $V_{\rm rel}$ is the relative
velocity of approach of the binaries at infinity,
\begin{equation}
S_{\rm bb} \, \sim \, {2\pi G(m_1 + m_2)a \over V_{\rm rel} ^2 }
\label{sigma}
\end{equation}
is the gravitationally focused cross section (Leonard 1989), and $a$
is the binary semimajor axis. For the sake of simplicity, we
consider the equal-mass and equal-energy binaries, and assume that
one of the binary components is a $40\, \msun$ main-sequence star while
its companion is a He core of mass of $5\,\msun$. From Eqs.\,(\ref{rate})
and (\ref{sigma}), one has the mean time between binary-binary
collisions (cf. Leonard 1989)
\begin{eqnarray}
t_{\rm bb} \sim 8\times 10^3 \, {\rm yr} \, \left({r_{\rm c} \over
0.01 \, {\rm pc} }\right)^{-3} \left({n_{\rm b} \over 10^7 \, {\rm
pc}^{-3} }\right)^{-2} \ \nonumber \\
\times \left({V_{\rm rel} \over 5 \, {\rm km} \, {\rm
s}^{-1} }\right) \left({m_1 \over 40 \msun
}\right)^{-1} \left({a \over 50\rsun}\right)^{-1} .
\label{bb}
\end{eqnarray}
It follows from Eq.\,(\ref{bb}) that the collisional time-scale is
short enough (i.e. $\la 10^4$ yr) if more than 10 massive
binaries exist in the core of radius $\sim 0.01$ pc. Assuming that
almost all stars more massive than $\sim 20\msun$ are segregated to
the cluster core and exchanged in binaries, one has the minimum mass
of the YMSC of $\sim 10^4 \msun$.

The recent discovery (Muno et al. 2006) of an anomalous X-ray pulsar
in the $(4\pm 1)\times 10^6$-yr old massive ($\ga 10^5 \msun$)
Galactic star cluster Westerlund\,1 (Clark et al. 2005) implies that
the progenitor of this neutron star was a star with a zero-age
main-sequence mass $\ga 40\msun$ (cf. Vanbeveren, Van Rensbergen \&
De Loore 1998). Therefore, one can expect that the hypervelocity
helium cores of massive stars ejected from the cores of YMSCs during
the first several Myr of cluster evolution will end their
lives as hyperfast neutron stars (pulsars).

\subsection{Exchange encounters between binary stars and an IMBH}
\label{subsec:nb}

Let us assume that the core of a YMSC harbours an IMBH of mass
$M_{\rm BH} =100-1\,000\msun$, formed either through a runaway sequence
of mergers or from the core collapse of a massive population
III star (see Sect.\,\ref{sec:ymsc}).

A close encounter with the IMBH results in the tidal breakup of the
binary, after which one of the binary components becomes bound to
the IMBH while the second one (usually the least massive star) is
ejected with a high speed, given by (Hills 1988; Yu \& Tremaine
2003):
\begin{equation}
V_{\infty} \sim \left({GM_{\rm BH} \over r_{\rm t} }
\right) ^{1/4} \left[{4Gm_1 ^2 \over a(m_1 +m_2)} \right]^{1/4} \, ,
\label{hills-1}
\end{equation}
where
\begin{equation}
r_{\rm t} \sim \left( {M_{\rm BH} \over
m_1 + m_2 } \right) ^{1/3} a \label{tid}
\end{equation}
is the tidal radius and $m_1$ and $m_2$ are, respectively, the masses
of the bound and ejected stars ($m_1 >m_2$). The fastest
stars produced in this exchange process come from encounters involving
the tightest binary systems (e.g. tidal binaries), since the tighter
the binary the closer it can approach to the IMBH before tidal
breakup. Combining Eqs.\ (\ref{hills-1}) and (\ref{tid}), one has
\begin{equation}
V_{\infty} \sim \left( {M_{\rm BH} \over m_1 +m_2 }
\right)^{1/6} \left( {2Gm_1 \over a} \right)^{1/2} \, .
\label{hills-2}
\end{equation}

For a binary having a massive main-sequence component of mass $m_1
\gg m_2$ and $a \simeq 2.5r_{1,{\rm MS}}$ (cf. Statler, Ostriker
\& Cohn 1987), one has from Eqs.\ (\ref{hills-2}) and (\ref{MS})
\begin{equation}
V_{\infty} \sim 440 \, {\rm km} \,{\rm s}^{-1} \left( {M_{\rm
BH} \over \msun} \right)^{1/6} \left( {m_1 \over \msun}
\right)^{-1/60} \, . \label{m1>>m2}
\end{equation}
For $M_{\rm BH} = 100-1\,000\msun$ and $m_1 =40-100\msun$, one has
from Eq.\ (\ref{m1>>m2}) that the low-mass binary component (e.g. the
helium core of a massive star or a B star) can be ejected with a speed
$V_{\infty} \sim 900 - 1\,300 \, {\rm km} \, {\rm s}^{-1}$. The
ejection velocity could be somewhat higher if the binary involved in
the encounter consists of two helium cores (cf. Sect.\,4.1). For
example, assuming that $m_{1,{\rm He}} = 10\msun$ and
$m_{2,{\rm He}} =5\msun$, and $a\simeq 5-10\rsun$, one has
$v_{\infty}  \simeq 850 -1\,760 \, {\rm km} \, {\rm s}^{-1}$. Note that
the weak dependence of $V_{\infty}$ on $M_{\rm BH}$ implies that the
hypervelocity helium cores can be produced by exchange encounters with
stellar mass (i.e. $\sim 20\msun$) BHs.

The mean collision time between binary stars and an IMBH is (e.g.
Binney \& Tremaine 1987)
\begin{eqnarray}
t_{\rm coll} \sim 2\times 10^5 \, {\rm yr} \, \left({n_{\rm b}
\over 10^7 \, {\rm pc}^{-3} }\right)^{-1} \left({V_{\rm rel} \over 5
\, {\rm km} \, {\rm s}^{-1} }\right) \ \nonumber \\
\times \left({M_{\rm BH} \over 100
\msun }\right)^{-1} \left({a \over 30\rsun }\right)^{-1}
\, . \label{coll}
\end{eqnarray}
It follows from Eq.\ (\ref{coll}) that during the several Myr
of cluster evolution one can expect at least a dozen of close
encounters involving binaries with $a\geq 30\rsun$. Here, we
assume a central density in the cluster which exceeds the
observed values. Although such high density may exist during core
collapse, we note that the well-studied YMSCs in the Galaxy are not
in this stage of their evolution.

\subsection{Encounters between single stars and a binary IMBH}
\label{subsec:bbh}

Numerical simulations by G\"{u}rkan, Fregeau \& Rasio (2006) showed
that runaway
collisions and mergers of massive stars in YMSCs with initial binary
fraction larger than $\sim 10$ per cent could result in the origin of two
VMSs. The subsequent supernova explosions of these VMSs produce two IMBH, which
ultimately exchange into a binary (G\"{u}rkan et al. 2006; cf. Fregeau
et al. 2006).

The IMBH binary (IMBHB) gradually hardens due to the interaction with
stars in the cluster's core. When the binary separation reduces to
$a\la a_{\rm h} \simeq G\mu /4\sigma ^2$, where $\mu = M_1 M_2 /(M_1 +
M_2 )$ and $M_1$ and $M_2$ are the component masses of the IMBHB ($M_1
>M_2 $), most of stars passing in the vicinity ($\sim a$) of the IMBHB
are expelled from the core at high velocity. The average ejection
speed attained by the escapers is (Yu \& Tremaine 2003)
\begin{equation}
\langle V_{\infty} \rangle \, \sim \, \sqrt{{3.2GM_1 M_2 \over (M_1
+ M_2) a}} \, . \label{v_ej}
\end{equation}
Some escapers, however, can reach much higher velocities. For an
IMBHB with mass ratio of $\sim 1$ the maximum ejection velocity is
\begin{equation}
V_{\infty} ^{\rm max} \sim 1.5 \, V_{\rm bin} \, , \label{tut}
\end{equation}
where $V_{\rm bin} =[G(M_1 +M_2 )/a]^{1/2}$ is the relative velocity
of the binary components if their orbits are circular (Tutukov \&
Fedorova 2005). It is clear that the smaller $a$ the larger $\langle
V_{\infty} \rangle$ and $V_{\infty} ^{\rm max}$. There are, however,
two constraints on the minimum value of $a$, which limit the maximum
possible ejection velocity.

First, $a$ should be sufficiently large to prevent the tidal breakup
of stars passing through the IMBHB, i.e. $a \ga [(M_1 /m)^{1/3} +
(M_2 /m)^{1/3} ]r$, where $r$ is the radius of the star. For $r$
given by Eq.\, (\ref{MS}), one has
\begin{equation}
a \ga 0.8\left[ \left({M_1 \over \msun} \right)^{1/3} +\left({M_2
\over \msun} \right)^{1/3} \right] \left({m\over \msun}
\right)^{11/30} \rsun \label{atid}
\end{equation}
(for the sake of simplicity we assume a circular binary orbit). It
follows from Eqs.\, (\ref{v_ej})-(\ref{atid}) that the smaller the
mass of the star the larger velocity it, in principle, can attain;
note also that inequality (\ref{atid}) sets an upper limit on the
mass of a main-sequence star that can be ejected by a shrinking
IMBHB. For a main-sequence star of mass $m=8\msun$ (the minimum mass
of single stars producing neutron stars) and assuming that $M_1 =
500 \msun$ and $M_2 = 300\msun$, one has from Eqs.\ (\ref{atid}),
(\ref{v_ej}) and (\ref{tut}) that $a\ga 30\rsun$, $\langle
V_{\infty} \rangle \sim 1950 \, {\rm km} \, {\rm s}^{-1}$ and
\begin{eqnarray}
V_{\infty} ^{\rm max} \simeq 3\,380 \, {\rm km} \, {\rm s}^{-1}
\left({\nu \over 0.625}\right)^{-1/2} \left({M_1 \over 500\msun
}\right)^{1/2}  \ \nonumber \\ \times \left({a \over 30\rsun
}\right)^{-1/2} \, , \label{v_max}
\end{eqnarray}
where $\nu =M_1 /(M_1+M_2 )$.

The maximum ejection velocity could be somewhat higher if one
considers encounters involving massive post-main-sequence stars,
either a blue supergiant or a bare helium core. In this case, a
massive star can approach the IMBHB much closer than a main-sequence
star. Although the blue supergiant star will lose its hydrogen
envelope due to the tidal stripping, its helium core can pass within
several $\rsun$ from one of the binary components without being
disintegrated by the tidal force. To estimate the maximum possible
velocity attained by the helium core, one should consider the second
constraint on the binary separation. It follows from the requirement
that the gravitational radiation time-scale of the shrinking IMBHB
(Peter 1964),
\begin{eqnarray}
t_{\rm GWR} \, \simeq \, 1\times 10^6 \, {\rm yr} \, \left({\nu
\over 0.625}\right) \left({M_1 \over 500\msun }\right)^{-2}  \ \nonumber \\
\times \left({M_2 \over 300\msun }\right)^{-1} \left({a \over
30\rsun }\right)^{4} \, , \label{GWR}
\end{eqnarray}
should be larger than the mean collision time,
\begin{eqnarray}
t_{\rm coll} \sim 2.5\times 10^4 \, {\rm yr} \, \left({n \over 10^7
\, {\rm pc}^{-3} }\right)^{-1} \left({V_{\rm rel} \over 5 \, {\rm
km} \, {\rm s}^{-1} }\right)  \ \nonumber \\
\times \left({M_1 +M_2 \over 800\msun }\right)^{-1} \left({a \over
30\rsun }\right)^{-1} \, , \label{colli}
\end{eqnarray}
that is, $a\ga 15\rsun$. From Eq.\ (\ref{v_max}), one has that a
helium core passing through the IMBHB just before the latter merges
due to the gravitational emission can attain a speed as large as
$\sim 4\,780 \kms$.

\section{N-body simulations of hypervelocity stars}
\label{sec:sim}

In this section, we perform numerical simulations of three-body
scatterings with IMBHs, both single and in binaries, in order to
obtain the velocity distributions for the ejected stars for the
mechanisms described in Sects.\,\ref{subsec:nb} and
\ref{subsec:bbh}. The simulations are carried out using the {\tt
sigma3} package, which is part of the STARLAB\footnote{\tt
http://www.manybody.org/manybody/starlab.html} software environment
(McMillan \& Hut 1996; Portegies Zwart et al. 2001b). For a detailed
description of the set up of the scattering experiments see
Gualandris et al. (2005).

\subsection{Exchange encounters between binary stars and an IMBH}
\label{subsec:sim:nb}

First, we focus on interactions in which a binary consisting of two
$8\msun$ main-sequence stars [with radii of 3.5$\rsun$, see
Eq.\,(\ref{eq:rms})] encounters a single IMBH of mass in the range
$100-1\,000\msun$. The relative velocity at infinity between the
IMBH and the centre of mass of the binary is set to $5\kms$, in
accordance with typical dispersion velocities in YMSCs.  The
semimajor axis is $a$ = 0.1\,au to represent the case of the
tightest binaries.
\begin{figure}
\begin{center}
\includegraphics[width=8cm]{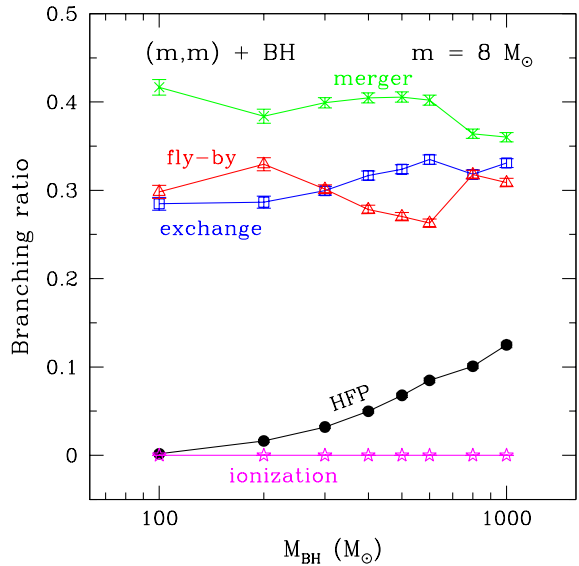}
\end{center}
\caption{Branching ratio for the outcome of encounters between a
  binary star and a single IMBH as a function of the IMBH mass.
  The two binary components are assumed to be 8$\msun$
  main-sequence stars and the semimajor axis is $a$ = 0.1\,au.  The
  different outcomes are: merger (crosses), fly-bys (triangles),
  ionization (stars), exchange (squares) and the subset of exchange
  results with a high-velocity escaper ($V_{\infty} \geq 700\kms$)
  (bullets).  The error bars represent the formal ($1\sigma$)
  Poissonian uncertainty of the measurement.}
\label{fig:branch}
\end{figure}
In Fig.\,\ref{fig:branch}, we present the probability of different
outcomes (branching ratios) as a function of the IMBH mass $\mbh$.
For each value of the IMBH mass, we perform a total of 2\,000
scattering experiments, which result either in a fly-by, an
exchange, or a merger of the components of the binary. Ionizations
never take place as the binary is too hard to be dissociated by the
IMBH (Heggie 1975). Mergers occur in a large fraction ($\sim$40 per
cent) of
encounters due to the small orbital period of the binary star and
mostly involve collisions between the binary components caused by
perturbations from the IMBH. Exchange interactions occur in about
30 per cent of encounters, with a probability increasing with the IMBH
mass.  Since the binary components have equal masses, the
probability of ejection for the two stars is equal in exchange
encounters. During such encounters, one of the main-sequence stars
is captured by the IMBH while the other star is ejected, possibly
with high velocity. These encounters are the relevant ones for the
production of hypervelocity stars.  If the velocity of the ejected
star exceeds 700$\kms$, we regard the star as a possible progenitor
of a hyperfast pulsar (indicated with HFP in the figure). The
figure shows that more massive IMBHs are more likely to eject stars
with hypervelocities.

\begin{figure}
\begin{center}
\includegraphics[width=8cm]{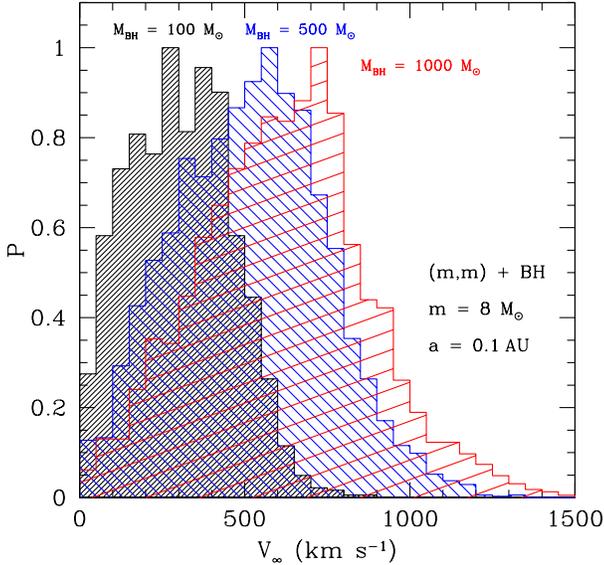}
\end{center}
\caption{Velocity distributions at infinity for escaping stars in encounters
  between an equal mass ($ m = 8\msun$) binary star and a single
  IMBH for different values of the IMBH mass:
  $\mbh=100\msun$ (left-hand panel), $500\msun$ (middle panel), $1\,000\msun$
  (right-hand panel). The binary semimajor axis is $a$ = 0.1\,au.}
\label{fig:vel}
\end{figure}
The distributions of velocities at infinity for the escaping stars are
shown in Fig.\,\ref{fig:vel} for three different values of the IMBH
mass: $\mbh=100\msun, 500\msun$ and $1\,000\msun$.  In order to obtain
stars with velocities $V_{\infty} \ga 700\kms$, an IMBH more massive
than a few hundred solar masses is required.  In the case of a
$500\msun$ IMBH, about 20 per cent of all exchanges result in an escape
velocity $\geq 700\kms$.  The fraction increases to 40 per cent for the
$1\,000\msun$ IMBH. We note that even an IMBH of mass $\mbh=100\msun$
can occasionally (in about 1 per cent of all exchanges) produce an escape
velocity $\geq 700\kms$.

In order to derive the probability of obtaining the largest possible
recoil velocities (see Sect.\,\ref{subsec:nb}), we perform
additional scattering experiments with the following parameters:
$m_1 = 40\msun$, $m_2 = 8\msun$ and $a = 0.15$\,au.
\begin{figure}
\begin{center}
\includegraphics[width=8cm]{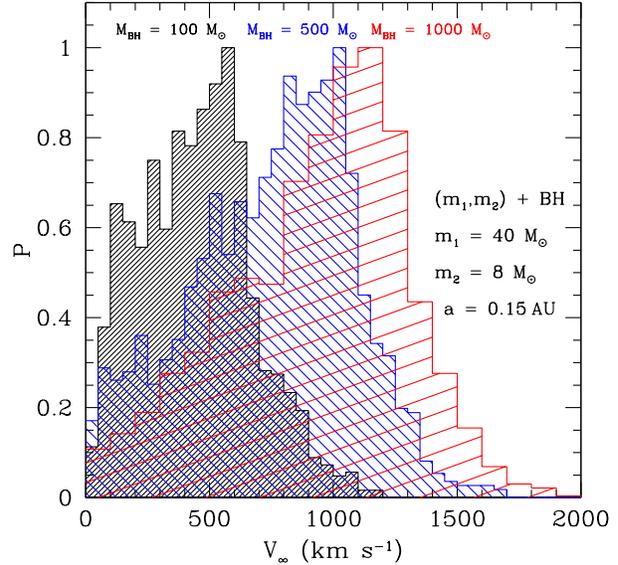}
\end{center}
\caption{Velocity distributions at infinity for escaping stars in encounters
  between a binary consisting of a primary star with mass $m_1 =
  40\msun$ and a secondary star with mass $m_2 = 8\msun$, and a single
  IMBH of mass $\mbh=100\msun$ (left-hand panel), $500\msun$ (middle panel),
  $1\,000\msun$ (right-hand panel).  For this case of unequal-mass binaries, we
  consider as escapers only the least massive stars ($m_2$). The binary
  semimajor axis is $a$ = 0.15\,au.}
\label{fig:vel2}
\end{figure}
The velocity distributions for escapers are shown in
Fig.\,\ref{fig:vel2} for three different values of the IMBH mass:
$\mbh=100\msun, 500\msun$ and $1\,000\msun$.  The maximum velocity obtained
for each set of parameters is consistent with the predictions derived
in Sect.\,\ref{subsec:nb}. The fraction of encounters resulting in
velocities larger than $700\kms$ increases from about 7 per cent for
$\mbh=100\msun$ to about 70 per cent for $\mbh=1\,000\msun$.

\begin{figure}
\begin{center}
\includegraphics[width=8cm]{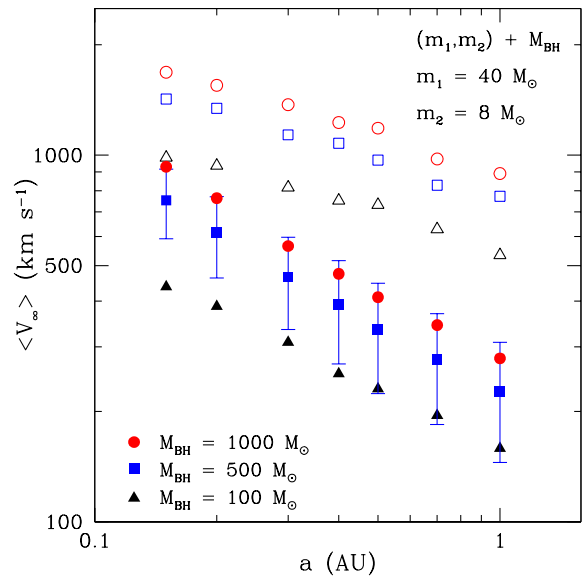}
\end{center}
\caption{Average recoil velocity of escapers as a function of the
  initial binary semimajor axis in the interaction of a binary
  star with IMBHs of different mass: $\mbh =100\msun$ (triangles),
  $\mbh =500\msun$ (squares), $\mbh =1\,000\msun$ (circles). Solid
  symbols represent the average velocity obtained from a set of 2000
  scattering experiments while the empty symbols indicate the velocity
  $V_{\rm max}$ for which 1 per cent of the encounters have $V_{\infty} >
  V_{\rm max}$. The errorbars indicate the 1$\sigma$ deviation from
  the mean. For clarity, we only show them for one data set.}
\label{fig:velo}
\end{figure}

The average velocity of escapers scales as $a^{-1/2}$, as can be seen
in Fig.\,\ref{fig:velo} and as is expected from Eq.\,(\ref{hills-2}).
The figure shows the average recoil velocity of escapers (solid
symbols) as a function of the initial binary semimajor axis for three
different values of the IMBH mass $\mbh=100\msun , 500\msun$ and
$1\,000\msun$.  The
empty symbols indicate the velocity $V_{\rm max}$ for which 1 per cent
of the encounters have $V_{\infty} > V_{\rm max}$. The average and the
maximum velocities increase with the mass of the IMBH,
as expected from energetic arguments.

\begin{figure}
\begin{center}
\includegraphics[width=8cm]{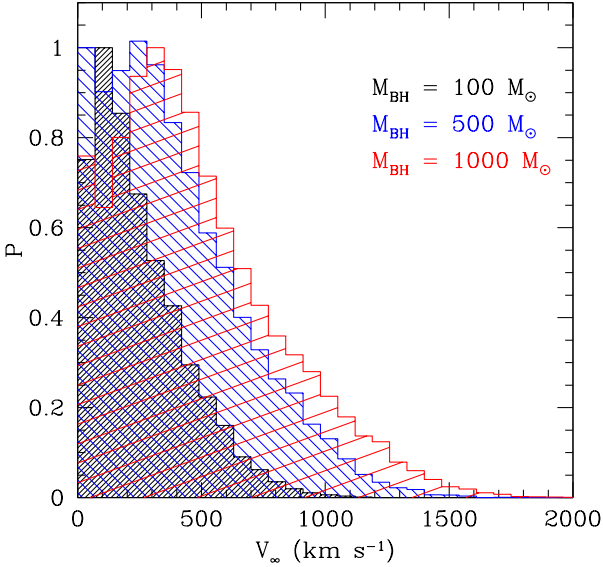}
\end{center}
\caption{Velocity distributions at infinity for ejected stars after an
         interaction between a $(40,8)\msun$ binary and an IMBH of mass
         $\mbh = 100\msun$ (left-hand panel), $500\msun$ (middle panel)
         and $1\,000\msun$ (right-hand panel). These velocity
         distributions are integrated over the
         entire range of orbital separations for the initial binary.
         Only secondary stars ($m_2 = 8\msun$) are considered as
         escapers in this figure.}
\label{fig:loga}
\end{figure}

In our systematic study of the effect of the initial semimajor axis
of the interacting binary, we performed further scattering experiments
adopting a homogeneous sampling in $\log a$.  If the distribution of
orbital separations in a star cluster is flat in $\log a$, like in the
case of young star clusters (Kouwenhoven et al. 2005), we can superpose
the results of these experiments in order to obtain the total velocity
distributions of escapers.  The resulting distributions for three
different values of the IMBH mass are presented in
Fig.\,\ref{fig:loga}.
The average and the maximum recoil velocities increase with the IMBH
mass, as in previous cases. The distributions appear much broader
than those in Fig.\,\ref{fig:vel2}, as a result of the sampling in the
semimajor axis.

In order to validate the theoretical predictions in
Sect.\,\ref{subsec:nb}, we simulated encounters between an IMBH and
binaries consisting of a main-sequence star and a He core.  We
considered main-sequence stars of mass $m_1 = 40\msun$ and
radius $r_1 = 11\rsun$
and He cores of mass $m_2 = 5\msun$ and radius $r_2 = 0.6\rsun$ [see
Eq.\,(\ref{eq:rhe})] in binaries of semimajor axis in the range $0.15 -
1.0$\,au.
\begin{figure}
\begin{center}
\includegraphics[width=8cm]{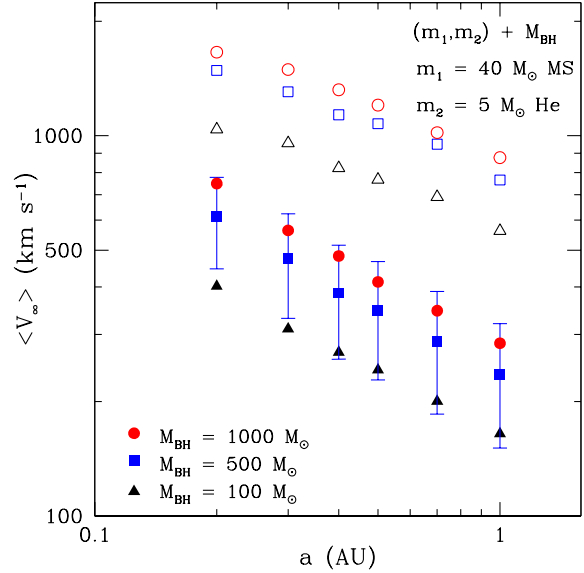}
\end{center}
\caption{Average recoil velocity of escapers as a function of the
  initial binary semimajor axis in the interaction between a binary
  consisting of a main sequence star with mass $m_1 = 40\msun$ and a
  He core with mass $m_2 = 5\msun$, and a single IMBH of mass
  $\mbh=100\msun$ (triangles), $500\msun$ (squares), $1\,000\msun$
  (circles). We consider as escapers only the least massive stars,
  i.e. the He cores.  Solid symbols represent the average velocity
  obtained from a set of 2\,000 scattering experimens while the empty
  symbols indicate the velocity $V_{\rm max}$ for which 1 per cent of the
  encounters have $V_{\infty} > V_{\rm max}$.  The errorbars indicate
  the 1$\sigma$ deviation from the mean. For clarity, we only show
  them for one data set.}
\label{fig:velohc}
\end{figure}
Fig.\,\ref{fig:velohc} shows the average recoil velocity of escapers
(solid symbols) as a function of the initial binary semimajor axis
for three different values of the IMBH mass
$\mbh=100\msun , 500\msun$, and $1\,000\msun$.
The empty symbols indicate the velocity
$V_{\rm max}$ for which 1 per cent of the encounters have $V_{\infty} >
V_{\rm max}$.  As in the case of encounters between binary stars
and a single IMBH, the average and the maximum velocities increase
with the mass of the IMBH. Maximum velocities are somewhat
higher compared to the case of two main-sequence stars (see
Fig.\,\ref{fig:velo}). This is due to the fact that, for a fixed
initial semimajor axis, a He star can get much closer to a BH
than a main-sequence star. The fraction of very close encounters,
however, is small and the average velocity of escapers is not
substantially higher than in the case of main-sequence binaries.

\subsection{Encounters between single stars and an IMBHB}
\label{subsec:sim:bbh}

Another exciting possibility, discussed in Sect.\,\ref{subsec:bbh},
is mediated by an encounter between a single star and an IMBHB.  We
simulate these encounters by considering a single star with a mass
of 8$\msun$ and a radius of 3.5$\rsun$ and binary components with
masses in the range $100-500\msun$, both in equal- and unequal-mass
binaries.  The semimajor axis is fixed to $a = 30\rsun \sim
0.14\,\rm au$ in all cases and the relative velocity at infinity
between the single star and the centre of mass of the IMBHB is set
to $5\kms$, as in the previous case.  We perform 5\,000 scattering
experiments for each set of parameters.  Except for a few mergers
(about 1-2 per cent of all cases), the vast majority of the encounters
result in a fly-by. With the random choice of impact parameter
adopted in the code (see McMillan \& Hut 1996; Portegies Zwart
2001a), typical recoil velocities in such large distance encounters
are modest.  The distribution of recoil velocities is strongly
peaked at small velocities ($V_{\infty} < 100\kms$) but shows a long
tail towards large velocities ($V_{\infty}
> 700\kms$).  The ejection velocities of escapers depend sensitively
on how close the single star approaches the IMBHB during the
encounter. High-velocity ejections are realized only if the star
approaches the binary within a distance comparable to the binary
separation.  The maximum ejection velocity for this scenario is
achieved when the incoming star passes through the binary system,
very close to one of the BHs. The distance of closest approach is
limited only by the tidal radius in the gravitational field of the
black hole, which is roughly given by $r_t = (\mbh/m)^{1/3} r$. The
fraction of encounters resulting in velocities larger than $700\kms$
increases from about 8 per cent for BHs of $100\msun$ to 20 per cent
for BHs of $500\msun$.

In order to prove the critical dependance of the ejection velocity at
infinity of escapers on the distance of closest approach to the IMBHB,
we perform another set of simulations with zero impact
parameter.  The values of the BH masses adopted in each set of
simulations are reported in Table\,\ref{tab:bbh}, followed by the
average velocity at infinity for escapers, the velocity $V_{\rm max}$
for which 1 per cent of the encounters have $V_{\infty} > V_{\rm max}$ and
the percentage of encounters for which the escapers achieve a recoil
velocity larger than $700\kms$ computed over 5\,000 scattering
experiments for each set of parameters.
\begin{table}
  \caption{List of BH masses adopted in the scattering experiments
    with zero impact parameter, followed by the average velocity at
    infinity for escapers, the velocity $V_{\rm max}$ for which 1 per cent of
    the encounters have $V_{\infty} > V_{\rm max}$ and the percentage
    of encounters for which the escapers achieve a recoil velocity
    larger than $700\kms$.  The maximum velocities should be
    considered as lower limits, as the actual value depends on the
    random sampling of the initial conditions in the scattering
    experiments.  }
  \label{tab:bbh}
  \begin{center}
    \begin{tabular}{ccccc}
      \hline
      $M_1$ &  $M_2$ & $\langle V_{\infty} \rangle$ & $V_{\rm max} (1 \, {\rm per} \, {\rm cent})$ & \#\\
      ($\msun$) &  ($\msun$) & ($\kms$) & ($\kms$) & ({\rm per} \, {\rm cent})\\
      \hline
      100 &  100 &  820 & 2175 & 53\\
      200 &  100 &  950 & 2380 & 66\\
      200 &  200 & 1130 & 3070 & 75\\
      300 &  200 & 1260 & 3230 & 80\\
      300 &  300 & 1430 & 3875 & 83\\
      500 &  300 & 1610 & 4170 & 87\\
      500 &  500 & 1845 & 4800 & 91\\
      \hline
    \end{tabular}
  \end{center}
\end{table}
The results shown in Table\,\ref{tab:bbh} show that
very large velocities can be achieved in close encounters between a
single star and a massive BH binary.
Encounters with impact parameters close to zero are, none the less,
very rare and we conclude that interactions between a stellar
binary and a single massive BH are more likely
to result in the ejection of an hypervelocity star.

\section{Discussion}
\label{sec:disc}

In this paper, we explored the hypothesis that the hyperfast pulsars
could be the remnants of a symmetric supernova explosion of a high-velocity
massive star which attained its peculiar velocity (similar to that of the
pulsar) in the course of strong dynamical three- or four-body encounters
in the cores of YMSCs. We estimated the maximum velocities obtainable in
these encounters and found that they are comparable to those measured for
hypervelocity stars. We therefore argue that the origin of hypervelocity
stars could be associated with dynamical processes in the cores of YMSCs.
Such star clusters are expected to be much more common than hitherto assumed
(see Sect\,\ref{sec:ymsc}).

Hypervelocity stars could also originate from other star-forming
galaxies. The position in the sky, the age and the observed velocity
of the hypervelocity star HE\,0437$-$5439 suggests that it was ejected
from the Large Magellanic Cloud (LMC) rather than from the Milky-way
Galaxy (Edelmann et al. 2005). Recently, Gualandris \& Portegies
Zwart (2007) investigated this hypothesis by simulating interactions
between a binary and an IMBH in a young star cluster in the LMC and
found that the hypervelocity star HE\,0437$-$5439 could have been
accelerated by one of the young star clusters in the LMC, but this
would require an IMBH of $\apgt 1\,000\msun$.

The cores of globular clusters may also give rise to ejecting stars
and  millisecond pulsars with high speed, in particular, those globular
clusters which have a high central density and are rich in recycled
pulsars. Some globular clusters may even harbour an IMBH, which are
known to be able to produce hypervelocity stars, and those clusters
form a natural channel for production of solitary millisecond pulsars,
whose velocities, in principle, can reach very high values.

In Sect.\,\ref{sec:hfp} and Sect.\,\ref{sec:sim}, we show that
main-sequence O-type stars cannot be ejected from YMSCs with a peculiar
velocity sufficiently high to explain the origin of hyperfast neutron
stars (unless a $\ga 1\,000 \,\msun$ IMBH is present in the cluster
center), but lower mass main-sequence stars or stripped helium cores
could be accelerated to hypervelocities. We find also that dynamical
processes in the cores of YMSCs can produce stars moving with
velocities of $\sim 200-400 \, \kms$, which are typical of pulsars (e.g.
Hobbs et al. 2005) and the bound population of halo B-type stars
(Ramspeck, Heber \& Moehler 2001; Brown et al. 2007), and therefore can contribute
to the origin of pulsar velocities in addition to asymmetric supernova
explosions and disruption of binaries following supernova explosions).
But even in these cases the cross-section for producing such
high speeds is small and we expect that only a fraction of
high-velocity runaways have been accelerated by this mechanism. Nevertheless,
some of velocities observed for pulsars may be explained by the proposed
scenario. Even though we present the cross-sections for scattering processes
discussed in Sect.\,\ref{sec:hfp} and Sect.\,\ref{sec:sim}, it is still hard
to estimate the production rate for the proposed mechanism. The
main uncertainty at the moment is in the number and characteristics of
the population of young and massive star clusters.

We argue also that some fast moving neutron stars could be the
descendants of hypervelocity helium cores, which due to their short
lifetimes ($\aplt 10^6$ yr) can only have been ejected quite recently.
As a consequence, the neutron star should not have had time to travel a long
distance from its parent star cluster. If we assume that the $\sim 2$ Myr
old pulsar PSR B1508+55 is indeed the remnant of a hypervelocity star,
then the pulsar separation from the Galactic plane of $\sim 2.5$ kpc implies
that its progenitor was a helium core (cf. Gvaramadze 2007). The
hypervelocity main-sequence stars of mass $\ga 8 \, \msun$ ejected at large
angles to the Galactic plane end
their lives at distances $>10$ kpc and thereby contribute to a population
of halo and intergalactic supernovae.

The proper motion vectors of several pulsars located in the
high-pressure interiors of their associated SNRs show a trend (Ng \&
Romani 2004) toward alignment with the pulsar rotation axes
(inferred from the symmetry of toroidal nebulae surrounding the
pulsars). This trend could be understood if pulsars receive at birth
a kick directed along their rotation axes. The observed alignment,
however, is not perfect. It is clearly pronounced only for the Vela
pulsar\footnote{For an alternative explanation of this alignment see
Radhakrishnan \& Deshpande (2001) and Deshpande \& Radhakrishnan
(2007).}. A natural explanation of the misalignment (if one adopts
that {\it all} neutron stars are kicked along their rotation axes)
is that the supernova progenitor star had an arbitrarily oriented
peculiar velocity, comparable with or larger than the kick velocity
received by its descendant at birth. One possibility is that the
supernova progenitor was a member of a binary system and that the
pulsar peculiar velocity is in part due to the recoil received by
the system after the first supernova explosion. Another possibility
is that the progenitor star was ejected from the parent YMSC due to
the processes discussed in Sect.\,\ref{sec:hfp}. Thus, the hyperfast
pulsars produced by supernova explosion of hypervelocity stars
should not show any alignment between their spin axes and proper
motion vectors. This inference, however, would be difficult to prove
since the hyperfast pulsars leave rapidly from the confines of their
parent SNRs [on a time-scale of $\sim 1.4\times 10^4\, v_{1\,000}
^{-5/3} (E_{51} /n_{\rm ISM})^{1/3}$ yr, where $v_{1\,000}$ is the
pulsar peculiar velocity in units of $1\,000 \, \kms$, $E_{51}$ is the
supernova energy in units of $10^{51}$ erg, and $n_{\rm ISM}$ is the
number density of the ambient interstellar medium]. The only known
hyperfast neutron star associated with a SNR (RX J0822$-$4300 in Puppis\,A; Hui
\& Becker 2006; Winkler \& Petre 2007) is a radio-quiet object
(Gaensler, Bock \& Stappers 2000) that makes it impossible to infer
the direction of its rotation
axis. One can also infer the orientation of the pulsar rotation axis
using the radio polarization measurements (e.g. Deshpande,
Ramachandran \& Radhakrishnan 1999; Johnston et al. 2005). As
expected, the proper motion of the hyperfast pulsar PSR B1508+55
shows a large misalignment of $\sim 71^{\degr}$ (or $\sim
19^{\degr}$, depending on whether or not the linear polarization of
the pulsar radio emission is parallel or orthogonal to magnetic
field; see Chatterjee et al. 2005).

\section{Acknowledgements}
We are grateful to D.R.Gies (the referee) for useful suggestions and
comments. VVG is grateful to A.M.Cherepashchuk, J.M.Fregeau, P.Hut,
N.Ivanova and A.V.Tutukov for useful discussions and acknowledges support
from the Deutsche Forschungsgemeinschaft. AG is supported by grant NNX07AH15G
from NASA. SPZ acknowledges support from the Netherlands Organization for
Scientific Research (NWO under grant No. 643.200.503) and the Netherlands
Research School for Astronomy (NOVA).

\end{document}